\documentclass[preprint,pteplogo]{ptephy}%%%%%% to generate preprint number with ptep logo

\preprintnumber{XXXX-XXXX} %%% Insert preprint number here

\usepackage{graphics} %for dealing with figures.
\begin{document}

\title{An aerogel Cherenkov detector for multi-GeV photon detection with low sensitivity to neutrons}

\author{\name{Y.~Maeda}{1\ast},
\name{N.~Kawasaki}{1},
\name{T.~Masuda}{1}\thanks{Present address : Research Core for Extreme Quantum World, Okayama University, Okayama 700-8530, Japan.},
\name{H.~Morii}{1},
\name{D.~Naito}{1},
\name{Y.~Nakajima}{1}\thanks{Present address : Lawrence Berkeley National Laboratory, Berkeley, California 94720, USA},
\name{H.~Nanjo}{1},
\name{T.~Nomura}{2},
\name{N.~Sasao}{3},
\name{S.~Seki}{1},
\name{K.~Shiomi}{4}\thanks{Present address : Institute of Particle and Nuclear Studies, High Energy Accelerator Research Organization (KEK), Tsukuba, Ibaraki 305-0801, Japan.},
\name{T.~Sumida}{1},
and \name{Y.~Tajima}{5}}
%%%%%%%%%%% The \name command should be used as \name{Insert author name here}{Insert affiliation number here}
%%%%% Please use \thanks for contributed author details

%%%%%%%%%%% The \affil command should be used as \affil{Insert affiliation number here}{Insert author address here}
\address{\affil{1}{Department of Physics, Kyoto University, Kyoto 606-8502, Japan}
\affil{2}{Institute of Particle and Nuclear Studies, High Energy Accelerator Research Organization (KEK), Tsukuba, Ibaraki 305-0801, Japan.}
\affil{3}{Research Core for Extreme Quantum World, Okayama University, Okayama 700-8530, Japan}
\affil{4}{Department of Physics, Osaka University, Toyonaka, Osaka 560-0045, Japan}
\affil{5}{Department of Physics, Yamagata University, Yamagata 990-8560, Japan}
\email{maeda\_y@scphys.kyoto-u.ac.jp}}

\begin{abstract}%
We describe a novel photon detector which operates under an intense flux of neutrons.
It is composed of lead-aerogel sandwich counter modules.
Its salient features are high photon detection efficiency and blindness to neutrons.
As a result of Monte Carlo (MC) simulations,
the efficiency for photons with the energy larger than 1~GeV is expected to be higher than 99.5\%
and that for 2~GeV/$c$ neutrons less than 1\%.
The performance on the photon detection under such a large flux of neutrons was measured for a part of the detector.
It was confirmed that the efficiency to photons with the energy \textgreater 1~GeV
was consistent with the MC expectation within an 8.2\% uncertainty.
\end{abstract}

\subjectindex{xxxx, xxx}

\maketitle

%\tableofcontents
%\linenumbers

%------------------------- Introduction ---------------------------%
\section{Introduction} \label{Sec_Intro}
%------------------------- Introduction ---------------------------%
An electromagnetic sampling calorimeter is one of the most popular detectors for detecting photons in high energy physics.
It is usually composed of alternate layers of high-Z converters and active materials sensitive to electrons and positrons.
Incident photons produce electromagnetic showers in the converter
and their energies are determined in the active layers.

\begin{figure}[t]
	\begin{center}
		\includegraphics[width=1\textwidth,bb=0 0 698 183]{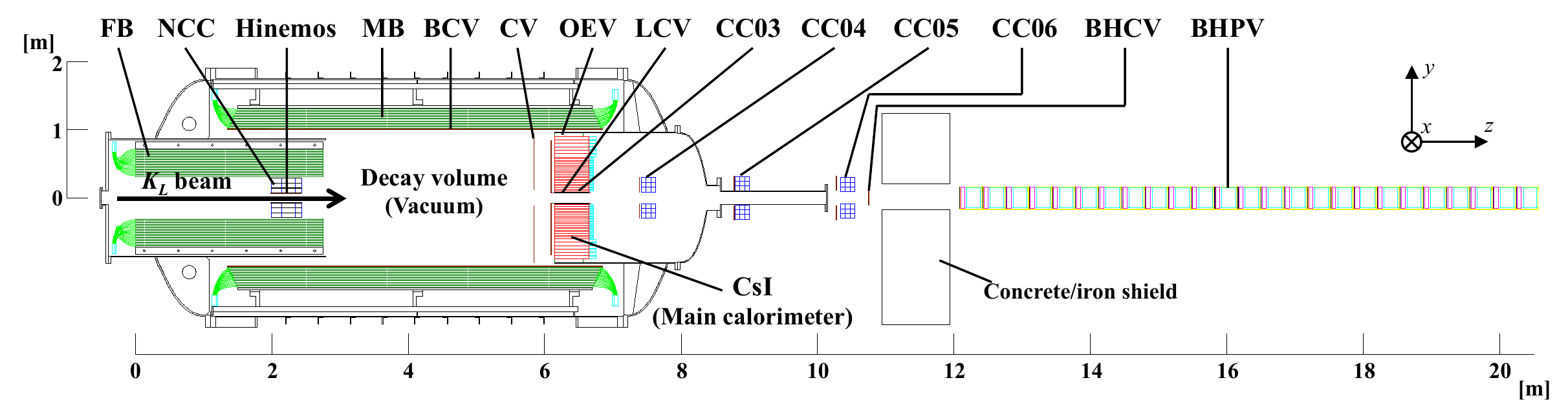}
        \end{center}
	\caption{Schematic cross-sectional view of the KOTO detector assembly.}
	\label{fig:KOTO}
\end{figure}

In this paper, we describe a novel photon detector with lead-aerogel sandwich used under an intense flux of neutrons.
We choose aerogel Cherenkov radiation as an $e^{\pm}$ sensing process
so that it is insensitive to heavier particles.
The detector is being used in a rare $K_{L}$ decay experiment\cite{KOTO-poposal,KOTO-review}
at the J-PARC Main Ring (MR)\cite{JPARC30}.
The experiment, named KOTO, aims to observe the CP-violating decay mode of $K_{L}\to \pi^{0} \nu \bar{\nu}$
with a sensitivity exceeding the standard model prediction ($Br \simeq (2.4 \pm 0.4) \times 10^{-11}$\cite{SMPrediction}).
An intense neutral kaon beam is needed to achieve the sensitivity.
The beam contains large amounts of photons and neutrons as well as $K_{L}$s
with typical energies of 10~MeV, 1.4~GeV and 2~GeV, respectively. %derived from average value (roughly)
Rates of the photons and neutrons are expected to be around 600~MHz each with the designed beam condition.
Figure~\ref{fig:KOTO} shows a cross-sectional view of the KOTO detector assembly. 
The signature of the $K_{L}\to \pi^{0} \nu \bar{\nu}$ decay is a pair of photons from $\pi^{0}$ and no other visible particles.
The two photons from the $\pi^{0}$ are detected with an electromagnetic calorimeter (CsI in Fig.~\ref{fig:KOTO})
placed downstream of an evacuated decay volume.
A number of veto counters surrounding the decay volume hermitically ensure the existence of no other visible particles at the same time.
The major background is expected to come from the decay $K_{L}\to 2\pi^{0}\to 4\gamma$,
in which two out of the four photons escape detection.
In order to suppress this type of background, photon detection with high efficiency is essential.
This requirement is also true for photons escaping into the beam direction,
and the energy of these photons ranges 100~MeV-5~GeV.
Thus, an efficient photon detector which works inside the beam with a large flux of neutrons is needed.
The crucial feature for the photon detector used in such an intense beam is blindness to neutrons,
which is to reduce single counting rates and overveto probabilities of the signal events.
Various requirements on the detector, including the next two, were evaluated by Monte Carlo (MC) simulations:
\begin{itemize}
 \item The efficiency is \textgreater99.5\% for photons with the energy of 1~GeV or greater.
 \item The efficiency is \textless1\% for neutrons with the momentum of 2~GeV/$c$.
\end{itemize}
In Sec.~\ref{sec_performance}, we report various design studies to satisfy these requirements,
including results of test experiments with positron and proton beams
for verification of the detector response and tuning of the simulation.
In Sec.~3, the performance on photon detection in the neutral beam was evaluated
for the partially-installed detector in the KOTO experimental area.
We add an appendix to describe a measurement of the transmittance of aerogel radiators,
which provided important parameters concerning the light yield.

%++++ figures 1 +++++++++++++++++++++++++++++++++++++++
%++++ figures 1 +++++++++++++++++++++++++++++++++++++++

%------------------------- Design ---------------------------%
\section{Design and expected performance} \label{sec_performance}
%------------------------- Design ---------------------------%

In this section, we first explain the basic concepts of the detector and describe its components.
Next, results of two test experiments with positrons and protons are reported.
Finally, we present results of simulation studies on the performance of the designed detector.

%------------------------------%
\subsection{Basic design} \label{Sec_Design}
%------------------------------%

\begin{figure}[t]
	\begin{center}
		\includegraphics[width=\textwidth,bb=0 0 727 164]{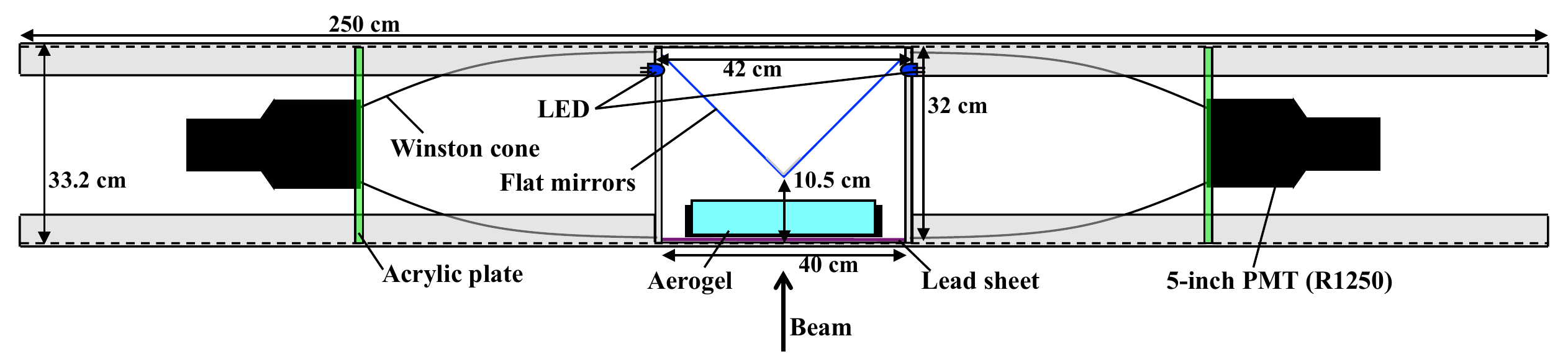}
        \end{center}
	\caption{Diagram of a single module (top view).}
	\label{fig:BHPV Single Module}
\end{figure}

The detector, named BHPV (Beam Hole Photon Veto),
is placed at the most downstream part of the KOTO detector assembly
as shown in Fig.~\ref{fig:KOTO}.
It is composed of 25 layers of modules along the beam.
The structure of a single module is shown in Fig.~\ref{fig:BHPV Single Module}.
The main part of the module consists of a lead converter and an aerogel radiator
as well as light collecting mirrors and photomultiplier tubes (PMTs).

%++++ figures 2 +++++++++++++++++++++++++++++++++++++++
%++++ figures 2 +++++++++++++++++++++++++++++++++++++++

%------------------%
\paragraph{Design concepts}
%------------------%
We chose a lead-aerogel sandwich counter as the module.
Photons are detected through Cherenkov radiation of the converted electrons and positrons in the aerogel,
which is known to have a small index of refraction ranging between 1.007-1.13. %values are taken from PDG2014 (p420)
This enables us to reduce sensitivity to neutrons
since they tend to produce slow particles which yield no or less Cherenkov light than $e^{\pm}$.

In order to achieve high photon efficiency, 
optimization of the converter and radiator thicknesses and the refractive index of the aerogel is important.
In general, a large number of sampling is required
because each converter should be thin enough to reduce shower particles stopping inside the converter
and the total thickness should be large enough to ensure conversion of photons into showers.
In our case, the total converter thickness of 10 X$_{0}$ and 25 samplings are adopted.
The refractive index of aerogel is chosen as $n=1.03$ by optimizing the photon efficiency and neutron blindness.

The arrayed configuration along the beam has an additional merit of reducing neutron sensitivity.
We note that electromagnetic showers by high energy photons tend to develop into the forward direction
while secondary particles such as protons and pions produced 
by neutron interactions have more isotropic angular distributions.
Thus, by defining photons as the events with hits in three or more consecutive modules, 
we can remove neutron events substantially.
Contribution from photons with the energy smaller than 50~MeV in the beam can also be reduced by this requirement.
Quantitative results of studies, performed with MC simulations, can be found in Sec.~\ref{MCStudyWithFullModule}.  
In the following, we describe the structure of a single module in more detail.

%--------------------%
\paragraph{Structure of a single module}
%--------------------%
Each single module consists of a lead sheet and aerogel tiles followed by a light collection system and PMTs.
The thickness of the lead sheet is 1.5-3.0~mm.
Two types of aerogel tiles with different sizes and optical qualities are used.
They are named type-M and type-A as listed in Table~\ref{table:Aerogel}.
Several layers of type-M (type-A) tiles are stacked into $3 \times 3$ ($2 \times 2$) grid
in order to cover the transverse size of 300-mm-square,
oversizing the actual neutral beam of 200-mm-square to detect diverging photons from $K_{L}$ decays.
These tiles are wrapped with a thin polyvinylidene chloride sheet,
whose transmittance is 90\% for visible light, to maintain their rigidity.
%with a thin (11 $\mu$m thick with transmittance of 90\%) polyvinylidene chloride sheet to avoid falling down.
%Dry nitrogen is circulated inside the module to keep environment humidity free. 
The optical system has two identical arms, each of which consists of a flat mirror, 
a Winston cone\cite{Winston} for collecting light, and a 5-inch PMT.
The advantages of the dual readout system include efficient and uniform light collection.
In addition, single counting rates are cut in half,
alleviating possible performance deterioration under high rate operation.
The flat mirror is made of a 0.75-mm thick aluminum sheet coated by an anodizing method.
The reflectivity is 85\% over the visible light region.
The Winston cone (480-mm-long) is designed to funnel the Cherenkov light from the input aperture of 300~mm in diameter %481.2mm in detail
into the output aperture of 120~mm.
It is made of an aluminum sheet by deep-draw processing; 
its inner surface is coated with aluminum by vapor deposition.
The average reflectivity is 85\% for visible light.  
The 5-inch PMT, Hamamastu R1250\cite{Hamamatsu}, has a bialkali photocathode with borosilicate glass.
Its quantum efficiency peaks around the wavelength of 400~nm with value of 20\%, 
according to its catalog information.
Light emitting diodes (LEDs) are installed for calibration of the PMTs.

%---------------------Table-------------------------%
\begin{table}[t]
\begin{center}
\caption{Parameters of the aerogel radiators.
	Type-M tiles were used in the calibration measurement in Sec.~\ref{PositronBeamTest}
	and type-A in the simulation study (Sec.~\ref{MCStudyWithFullModule})
	and the physics run (Sec.~\ref{PhysicsRun}).}
\begin{tabular}{ccccc}
\hline\hline
Type    & Refractive  &  Dimensions  & Configuration &  Transmission length [cm]   \\ 
        &  index ($n$)&   [mm$^3$]  & of stacking & (at the wavelength of 400 nm)  \\ \hline
M & 1.03 & $100 \times 100 \times 11 $  & $3\times3$ & 5.07 \\ 
A & 1.03 & $159 \times 159 \times 29 $  &  $2\times2$ & 3.35 \\ \hline\hline
\end{tabular}
\label{table:Aerogel}
\end{center}
Note: The transmission length, defined as the path length at which the original intensity is reduced to $1/e$, 
is calculated with Eq.~(\ref{eq:transmittance}) using the measured parameters.
See Appendix for detail.
\end{table}
%---------------------Table-------------------------%

\begin{figure}[t]
	\begin{center}
		\includegraphics[width=1\textwidth,bb=0 0 639 166]{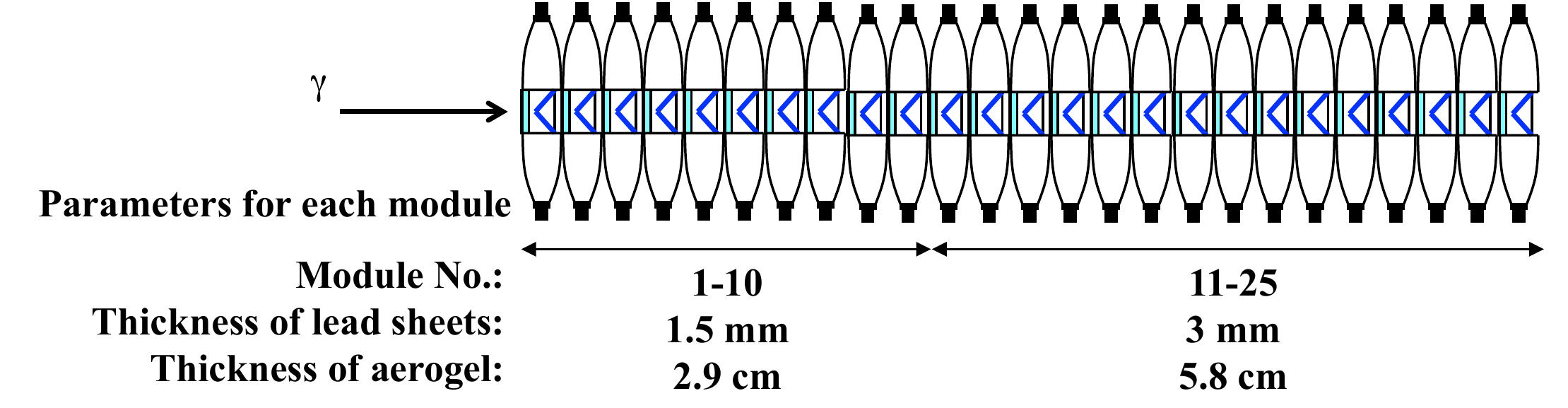}
        \end{center}
	\caption{Layout of the BHPV detector.}
	\label{fig:BHPV Layout}
\end{figure}

%--------------------%
\paragraph{Arrayed configuration of modules}
%--------------------%
Twenty-five modules are arranged along the beam axis.
Thickness of the lead and aerogel radiator for each module are shown in Fig.~\ref{fig:BHPV Layout}.
This configuration, used in the simulation studies in Sec.~\ref{MCStudyWithFullModule},
is referred as the reference configuration.
They are optimized according to experimental conditions such as beam intensity
in order to keep the photon detection efficiency high and the single counting rates as low as possible.
For example, 
thinner lead sheets and aerogel in the upstream modules, as in the reference configuration,
help to reduce the counting rates in these modules where high rates are expected.

%++++ figures 3 +++++++++++++++++++++++++++++++++++++++
%++++ figures 3 +++++++++++++++++++++++++++++++++++++++

%------------------------------%
\subsection{Photoelectron yield measurement with a positron beam} \label{PositronBeamTest}
\begin{figure}[h]
	\begin{center}
		\includegraphics[width=\textwidth,clip,trim=20 40 25 30]{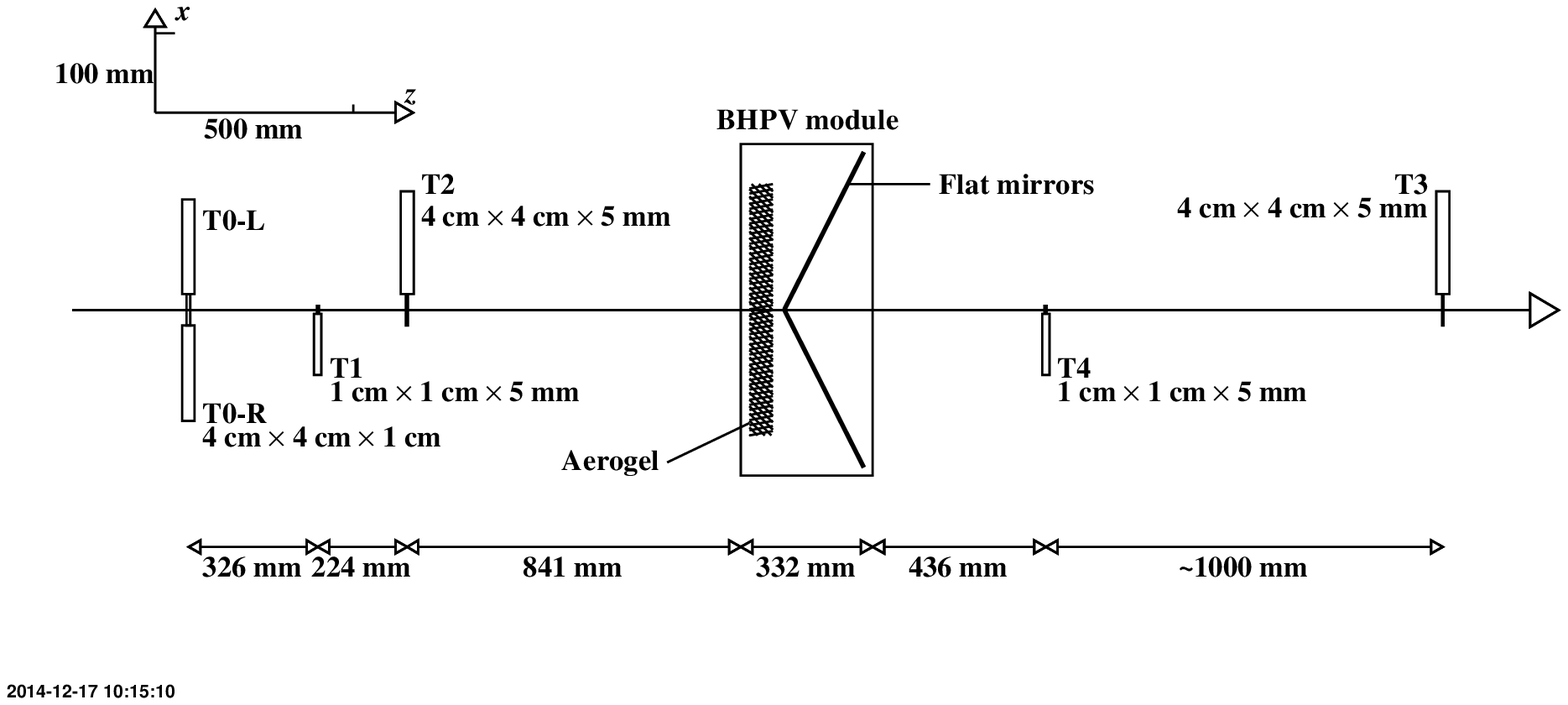}
		%trim = left bottom right top 
        \end{center}
	\caption{Layout of the photoelectron calibration measurements using positrons.
		$z$-axis indicates the beam direction.
		The name and size are shown for each trigger counter.
		Only a part of the BHPV detector module, an aerogel radiator and flat mirrors, is drawn for simplicity 
        (see Fig.~\ref{fig:BHPV Single Module} for the optical system).
        The lead converter was not inserted for this experiment.}
	\label{fig:Positron Beam Test Layout}
\end{figure}

%------------------------------%
The average number of observable photoelectrons (p.e.) produced by a single relativistic electron traveling through the aerogel radiator
is the most important quantity.
The value was obtained by the ``calibration experiment'' with a positron beam.
The experiment was performed using a 600~MeV/$c$ positron beam 
at the Laboratory of Nuclear Science\footnote{Presently the Research Center for Electron Photon Science} of Tohoku University, Japan in 2009.
Five layers of type-M aerogel tiles in Table~\ref{table:Aerogel} were used in this measurement.
Figure~\ref{fig:Positron Beam Test Layout} shows a schematic view of the experimental layout.
Plastic scintillation counters were placed in both the upstream and downstream of the detector module.
The trigger signal was formed essentially by the coincidence signals of T1 and T4 counters, 
both of which had dimensions of 1~cm $\times$ 1~cm in cross section and 0.5~cm in thickness.

%++++ figures 6 +++++++++++++++++++++++++++++++++++++++
%++++ figures 6 +++++++++++++++++++++++++++++++++++++++

%++++ figures 7 +++++++++++++++++++++++++++++++++++++++
\begin{figure}[h]
	\begin{center}
		\includegraphics[width=0.5\textwidth,clip,trim=10 20 55 35]{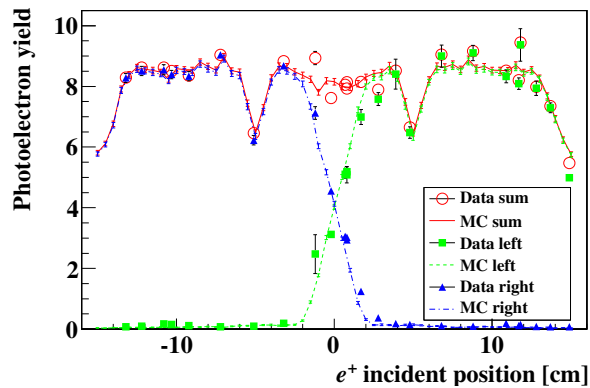}
		%trim = left bottom right top
        \end{center}
	\caption{The data (markers) and MC simulation results (lines)
	of the light yields as a function of the horizontal ($x$) position of the positron ($e^{+}$) beam.}
	\label{fig:Positron Beam Test Results}
\end{figure}
%++++ figures 7 +++++++++++++++++++++++++++++++++++++++

Figure~\ref{fig:Positron Beam Test Results} shows the photoelectron yields as a function of the horizontal beam position.
In the measurement, an output of each PMT was converted to the number of photoelectrons using an LED calibration data.
Results of corresponding MC simulations are shown in the same figure with lines,
where ray tracing was performed for individual Cherenkov light produced in the aerogel.
Here, we took into account various loss factors such as geometrical acceptance and reflectivity of the optical system,
quantum efficiency of the PMT, and measured transmittance of the aerogel tiles\footnote{
Detail of the measurement is in Appendix.}.
The dips around $x=\pm 5$ cm are due to boundaries between the aerogel tiles.
The simulation successfully reproduces the uniform photoelectron yield over the entire region in the data. 
The absolute scale of the simulation was corrected so that the average photoelectron yield agreed with that of the data.
This scale factor is referred to as the ``calibration factor'' and is found to be 0.55 in this measurement.
The origin of this correction is considered to be
due to uncertainty in the quantum efficiency of the PMT and deterioration of the aerogel surface during transportation and handling.
In fact, fine fragments produced by frictions of tiles were observed on the surface during the measurement.
It was likely that they caused an additional loss of the photoelectron yield.

%------------------------- performance tests ---------------------------%
\subsection{Measurement of hadronic response}  
%------------------------- performance tests ---------------------------%
In this section, we describe test results with a proton beam.
The purposes are to validate the MC simulation for hadronic interactions
and to examine experimentally the detector response to hadrons.

%++++ figures 8/9 +++++++++++++++++++++++++++++++++++++++
\begin{figure}[h]
\begin{minipage}{9.25 cm}
	\begin{center}
		\includegraphics[width=\textwidth,bb=0 0 460 253]{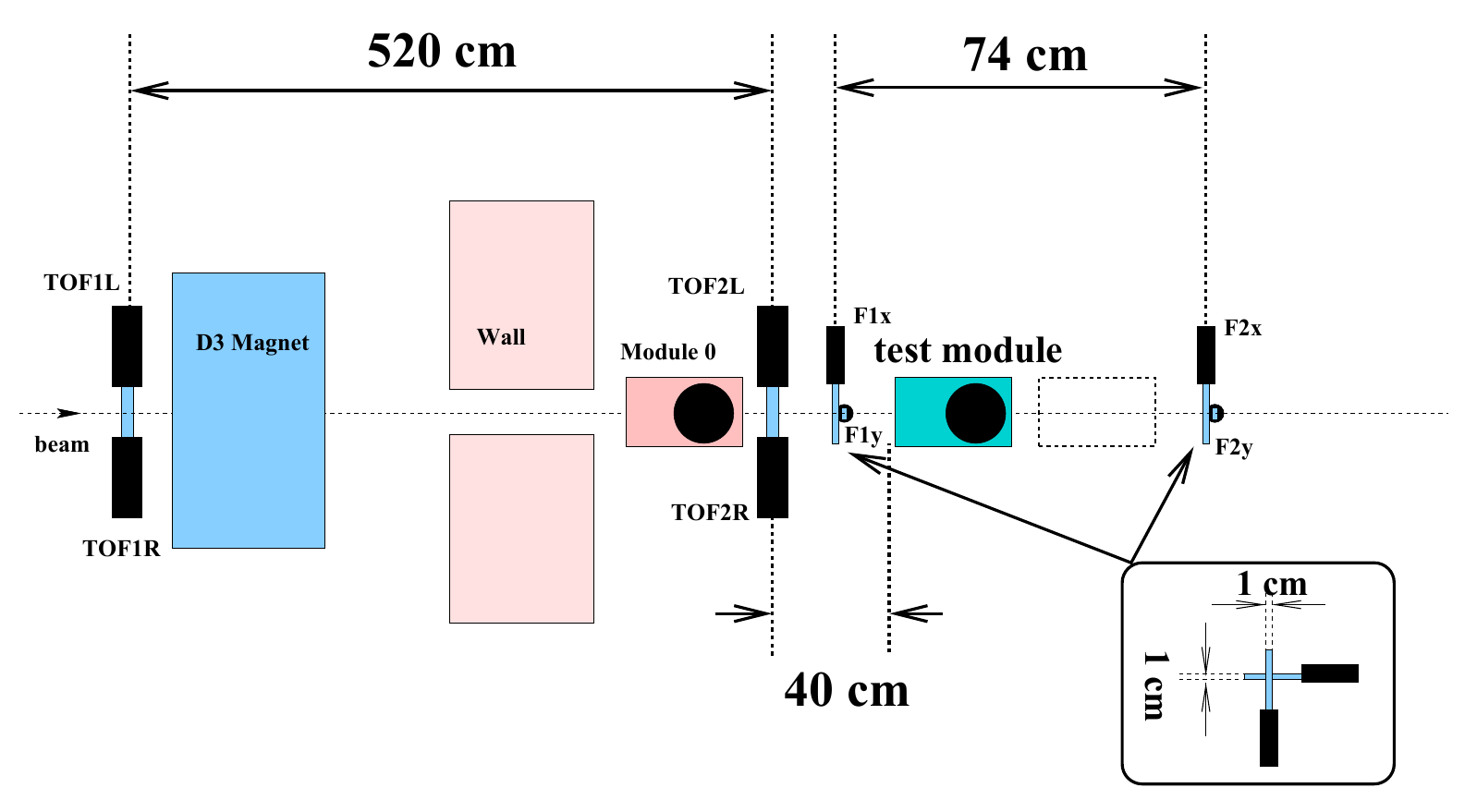}
        \end{center}
	\caption{Layout of the proton beam experiment. 
        The module under test was placed at the position labeled ``test module''.}
	\label{fig:Experimental setup done with a proton beam}
\end{minipage}
\begin{minipage}{0.25cm}$\;$\end{minipage}
\begin{minipage}{5.5 cm}
	\begin{center}
		\includegraphics[width=\textwidth,bb=0 0 269 250]{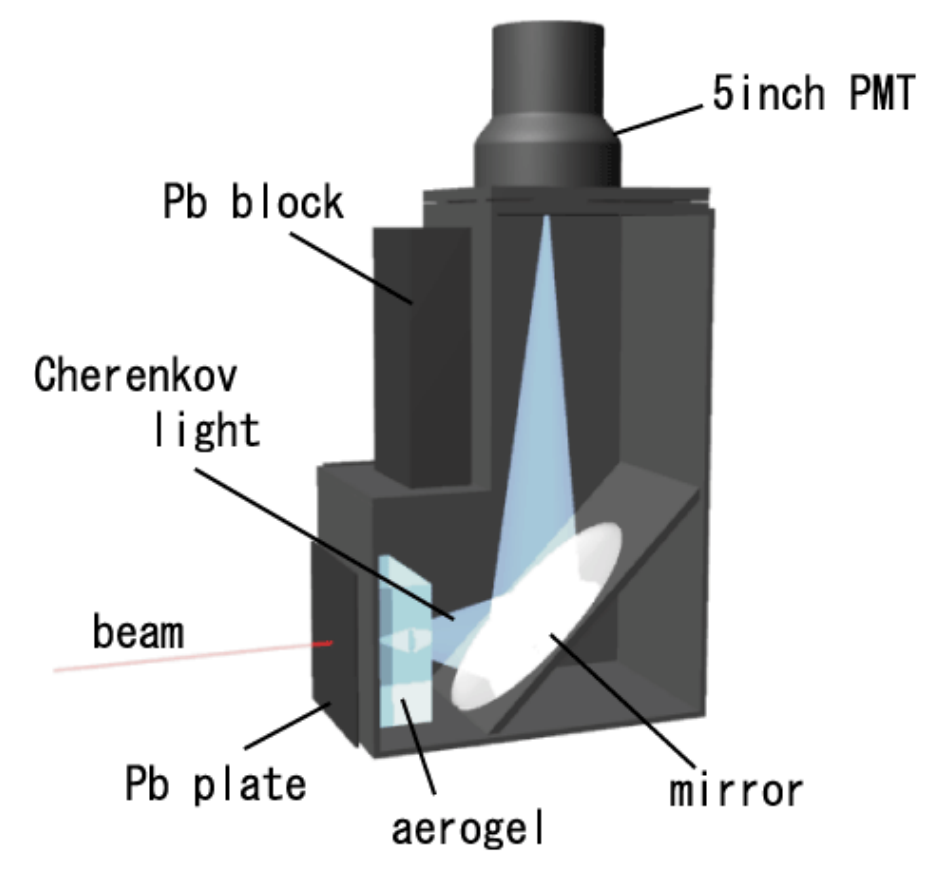}
        \end{center}
	\caption{Schematic view of the test module used in the proton beam experiment.}
	\label{fig:Aerogel module used in a proton beam}
\end{minipage}
\end{figure}
%++++ figures 8/9 +++++++++++++++++++++++++++++++++++++++

The experiment was performed at the 12~GeV Proton Synchrotron of High Energy Accelerator Research Organization (KEK), Japan in 2002.
Figure~\ref{fig:Experimental setup done with a proton beam} shows its schematic layout.
Figure~\ref{fig:Aerogel module used in a proton beam} shows the test module
with one 5-inch PMT and a parabola mirror as an optical system.
A 20-mm-thick lead converter and five aerogel tiles with the same dimensions as type-M\footnote{
Aerogel used in this measurement had different optical characteristics from both of type-M and type-A.
Transmittance and calibration factor were separately measured for this aerogel.}
were placed in the module. 
A much thicker converter than that in the reference configuration in Sec.~\ref{Sec_Design} was used to enhance hadronic interactions.
The trigger signal was formed by a coincidence signal from the time-of-flight counters (TOF1 and TOF2) and two   
1-cm-wide mutually orthogonal counters (F1x and F1y).
Particle identification was made by time-of-flight information measured by TOF1 and TOF2.

\begin{figure}[h]
	\begin{center}
		\includegraphics[width=0.5\textwidth,bb=0 0 212 198]{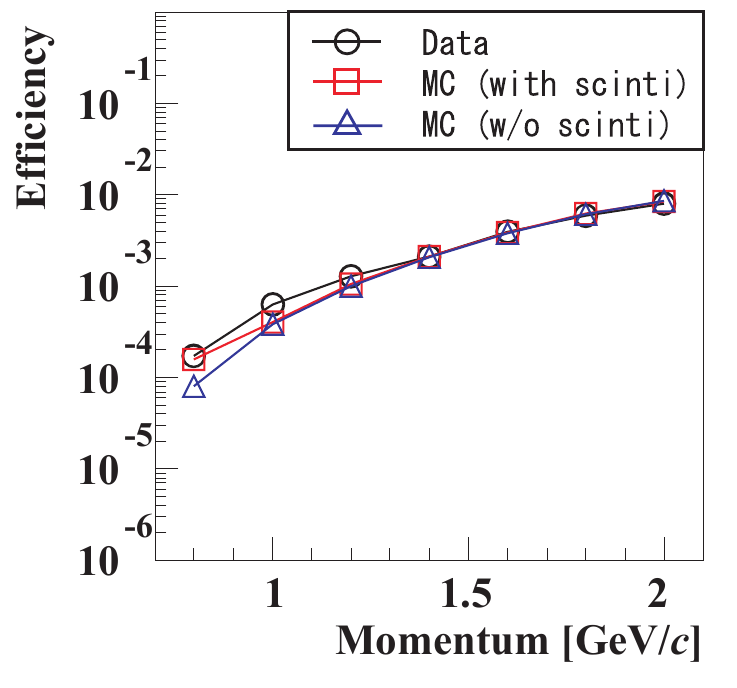}
        \end{center}
	\caption{Detection efficiency for protons as a function of the momentum.
       The experimental results using a 1.75~p.e. threshold are shown by black open circles together with
       corresponding MC simulation results with and without the contribution from the nitrogen gas scintillation
        by red open squares and blue open triangles, respectively.} 
	\label{fig:Hadron Beam Test Results}
\end{figure}

Figure~\ref{fig:Hadron Beam Test Results} shows the results of the measurement with MC expectation;
the efficiency with the threshold of 1.75~p.e. is plotted as a function of the proton momentum.
Note that the efficiency obtained in this measurement is
for a single test module.
In spite that a proton itself dose not generate Cherenkov light in this momentum range,
it can make a signal in the module through the generation of knock-on electrons
and secondary particles such as $\pi^{0}$s.
In addition, the scintillation light from nitrogen in the air also makes contribution\cite{Morii}.
The agreement between the data and MC in Fig.~\ref{fig:Hadron Beam Test Results} shows
that the response of this detector to protons is well-understood.
These results validate our MC simulations on the neutron blindness of the detector.

%++++ figures 10 +++++++++++++++++++++++++++++++++++++++
%++++ figures 10 +++++++++++++++++++++++++++++++++++++++

%------------------------------%
\subsection{Expected performance}	\label{MCStudyWithFullModule}
%------------------------------%
In this section, we present the expected performance studied by MC simulations.
We have adopted the GEANT4 simulation codes\cite{GEANT-4}.
We focus on the photon efficiency and neutron blindness.

\paragraph{Condition of the simulation}
The reference configuration described in Sec.~\ref{Sec_Design} was employed in the simulation study.
The type-A aerogel, which are used in the KOTO physics run, was assumed.
Transmittance and calibration factors, separately measured for this aerogel, were implemented.
The simulation procedure is as follows.
Photons with various energies were injected uniformly over the detector upstream surface of 250-mm-square.
When $e^{\pm}$ tracks in the electromagnetic showers traversed the aerogel radiator,
Cherenkov light was emitted and the rays were traced
from the radiator to the PMT cathode.
The amount of the Cherenkov light at the PMT was converted to the number of photoelectrons by using the calibration factor.
The same procedures were applied
until all the shower particles exited from the entire detector or lost their energy completely.

%++++ figures 4 +++++++++++++++++++++++++++++++++++++++
\begin{figure}[h]
\begin{minipage}[t]{7.5cm}
	\begin{center}
		%trim = left bottom right top 
		\includegraphics[width=\textwidth,clip,trim=20 25 55 35]{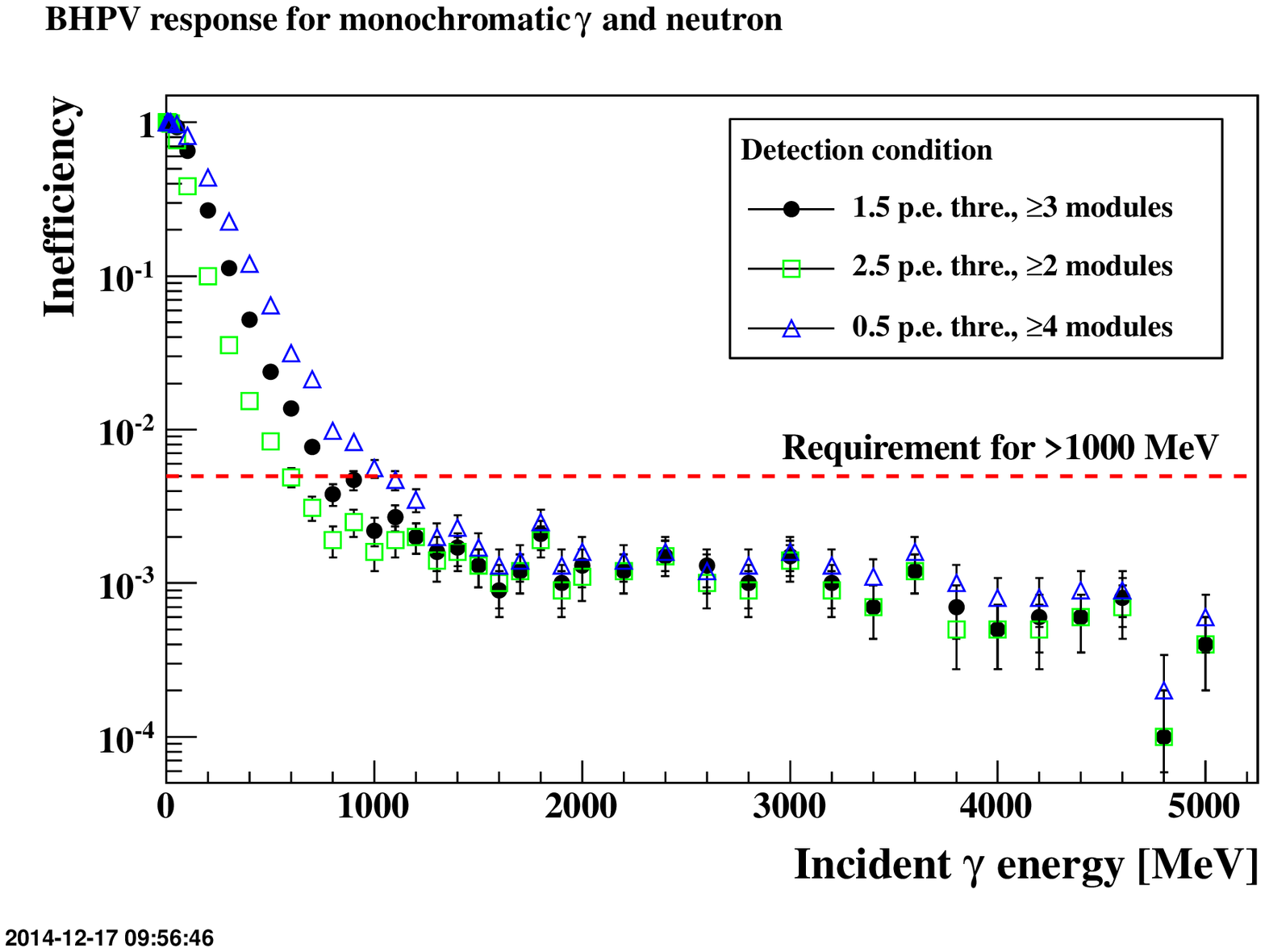}		
        \end{center}
	\caption{Photon inefficiency of the BHPV detector estimated by the MC simulations.
	The reference configuration is assumed.
        Three different definitions for the photon hits are shown; $\ge$3 consecutive modules with $>$1.5 p.e. (black solid circles), 
        $\ge$2 modules with $>$2.5 p.e. (green open squared), and $\ge$4 modules with $>$0.5 p.e. (blue open triangles).
        The red dashed line indicates the upper bound of the requirement described in Sec.~\ref{Sec_Intro}.}
	\label{fig:Photon efficiency MC}
\end{minipage}
%++++ figures 4 +++++++++++++++++++++++++++++++++++++++
\begin{minipage}{0.5cm}$\:$ \end{minipage}
%++++ figures 5 +++++++++++++++++++++++++++++++++++++++
\begin{minipage}[t]{7.5cm}
	\begin{center}
		\includegraphics[width=\textwidth,clip,trim=20 25 55 35]{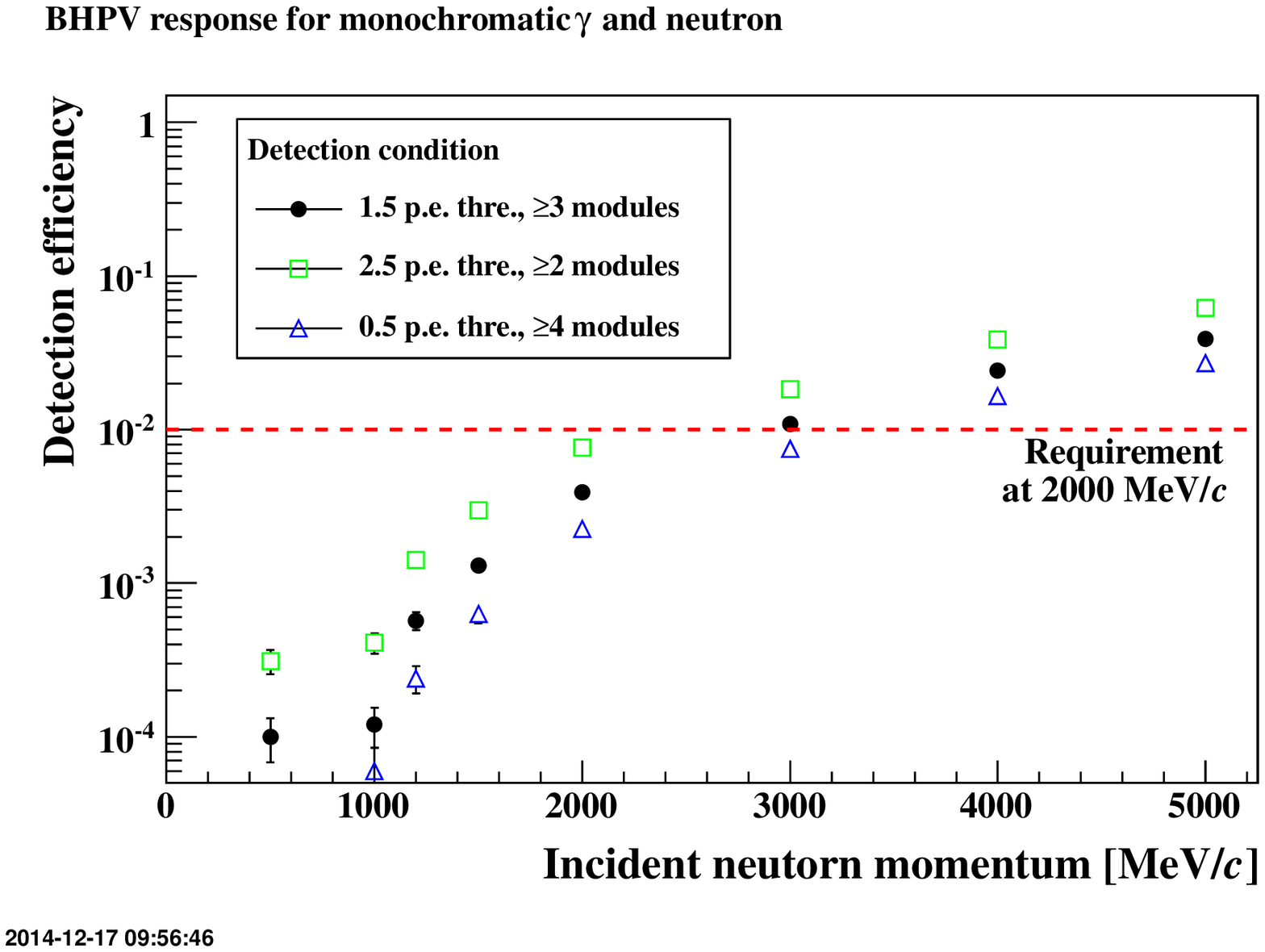}		
        \end{center}
	\caption{Neutron efficiency of the BHPV detector estimated by the MC simulations. The reference configuration is assumed.
		Meanings of the markers and the red dashed line are the same as in Fig.~\ref{fig:Photon efficiency MC}.
        }
	\label{fig:Neutron blindness MC}
\end{minipage}
\end{figure}
%++++ figures 5 +++++++++++++++++++++++++++++++++++++++

\paragraph{Photon efficiency}
The following algorithm was adopted to identify photons.
In a single module, a hit was recognized
when the output from either or both of the two PMTs exceeded a 1.5~p.e. threshold.
If three or more consecutive hits were recorded, then the event was identified as a photon.
Black solid circles in Fig.~\ref{fig:Photon efficiency MC} show the inefficiency as a function of the incident photon energy.
The simulation shows that the detector satisfies the photon efficiency requirement, $>$99.5\% for \textgreater1~GeV.

\paragraph{Neutron blindness}
Estimation of the neutron efficiency proceeded in the same way as the photon efficiency.
In this case, neutrons were injected, and hadronic showers were produced in the lead converter.
We used the hadron package of QGSP to simulate the neutron interactions.
All the charged particles were tracked and Cherenkov light was created
when the momentum was above the Cherenkov threshold.
The event identification algorithm was the same as in the photon case.
Figure~\ref{fig:Neutron blindness MC} shows the neutron efficiency as a function of incident neutron momentum.
The efficiency increases monotonically with the momentum,
and remains below 1\% for 2~GeV/$c$ neutrons,
satisfying the requirement specified in Sec.~\ref{Sec_Intro}.

Different algorithms may lead to different photon inefficiencies and neutron efficiencies.
Examples of such studies are shown in Fig.~\ref{fig:Photon efficiency MC}
and Fig.~\ref{fig:Neutron blindness MC}.
If the number of required consecutive hits is lowered, the efficiency for low energy photons increases,
but the efficiency for neutrons also increases.
These conditions can be optimized according to specific experimental situations.

%------------------------------%
\section{Photon identification in the neutral kaon beam}	\label{PhysicsRun}
%------------------------------%
As mentioned in Sec.~\ref{Sec_Intro},
a part of the entire BHPV detector was installed
in the experimental area together with other KOTO detectors.
This partial detector consists of 12 modules, 
and was loaded with 58-mm-thick type-A tiles and lead converters with different thickness:  
five 1.5-mm (No.1--5), five 3-mm  (No.6--10) and two with no plates (No.11--12)\footnote{
Since the identification of a photon signal requires hits in three or more consecutive modules,
lead in the last two modules dose not contribute to the total thickness effectively.
This is why the modules No.~11 and No.~12 do not have lead converters.}. 
Outputs from this detector were recorded by waveform digitizers of 500~MHz sampling,
which were custom-built for the KOTO experiment\cite{500MHzFADC}.
Multiple hits in a single counter were distinguished correctly even under the high-rate environment.
We present analysis results of the 100-hour data
obtained in the first physics data taking in May, 2013.
The beam power of the J-PARC MR was 24~kW,
which corresponds to the average neutron and photon rates of 100~MHz and 170~MHz\footnote{
The rates are with the kinetic energy larger than 1~MeV.},
respectively, from the MC simulation.
We focus on the detector response to high energy photons from $K_{L}$ decays
with the accompanied neutron and photon fluxes.

%------------------%
\subsection{Photon tagging with $K_{L}\to 3 \pi^{0}$ decay samples}
%------------------%

There were six photons in the $K_{L}\to 3 \pi^{0}$ decay.
In the analysis, we required five out of the six photons to hit the CsI calorimeter.
Kinematics of the decay allowed the reconstruction of the ``missing'' photon with a two-fold ambiguity.
This ``tagged photon'' technique was used to evaluate the performance of the BHPV with the collected data in the KOTO experiment.
We compared data with MC to validate the performance.
From the MC simulation, the ``missing photon,'' denoted as $\gamma_{6}$ below,
has an geometric acceptance of $\sim$3\%
in the direction of the BHPV.
Details of the CsI calorimeter can be found elsewhere\cite{KOTO-review, Masuda}.
We started with the selection of the five reconstructed photons.
For any two photons, when we assumed they were from $\pi^{0}$ decay,
the longitudinal vertex position was calculated:
\begin{equation}
M_{\pi^{0}}^{2} = 2e_{1}e_{2}(1-\cos\theta),
\label{eq:pi0Reconstruction}
\end{equation}
where $M_{\pi^{0}}$ is the $\pi^{0}$ mass, $\theta$ is the opening angle,
and $e_{1}$, $e_{2}$ were the photon energies.
We further assumed the transverse position of the $\pi^{0}$ to be at the beam-line.
Out of the five photons, there were 15 possible combinations to reconstruct two $\pi^{0}$ decays.
For each of these 15 combinations, there were two vertices.
We chose the correct combination by requiring the two vertices to be the same (best fit)
so that it is the common vertex of the $K_{L}$ decay\footnote{
We calculated $\chi_{z}^2=\sum_{i} (z_{i}-\bar{z})^2/\sigma_{i}^2$ for each combination, 
and chose the smallest one. Here $i$ denotes each $\pi^{0}$ candidate, $z_{i}$ ($\sigma_{i}$) is its vertex position 
(resolution), and $\bar{z}$ is the weighted-mean of the two vertex positions.}.
With the decay vertex known, momentum of the third $\pi^{0}$ was calculated.
We denoted $\gamma_{5}$ and $\gamma_{6}$ as the two photons from the third $\pi^{0}$:
\begin{eqnarray}
M_{\pi^{0}}^{2} & = & E_{3}^{2} - \sum_{i=x,y,z} P_{3,i}^{2} 	\nonumber \\
	& = & (e_{5}+e_{6})^{2} - \sum_{i=x,y,z} (p_{5,i}+p_{6,i})^{2}	\nonumber \\
	& = & (e_{5}+\sqrt{p_{6,x}^{2}+p_{6,y}^{2}+p_{6,z}^{2}})^{2} - \sum_{i=x,y,z} (p_{5,i}+p_{6,i})^{2},
	\label{Eq:Gamma5Rec}
\end{eqnarray}
where $E_{3}$, $e_{5}$, and $e_{6}$ were energies of the third $\pi^{0}$, $\gamma_{5}$ and $\gamma_{6}$, respectively.
$P_{3,i}$, $p_{5,i}$, and $p_{6,i}$ ($i=1,2,3$) are the $i$-th components of the momenta.
For the three unknowns of $\gamma_{6}$ momenta, the two transverse components ($p_{6,x}$, $p_{6,y}$) were determined
assuming that the parent $K_{L}$ has no transverse momentum. 
Equation~(\ref{Eq:Gamma5Rec}) is thereby quadratic for $p_{6,z}$.
For the two solutions of $p_{6,z}$, we obtained two $K_{L}$ invariant masses.
The solution with the larger (smaller) $p_{6,z}$ was called the ``forward''(``backward'') solution
and the corresponding $K_{L}$ mass was denoted as $M_{K_{L}}^{\mathrm{forward}}$($M_{K_{L}}^{\mathrm{backward}}$).

\begin{figure}%[h]
	\begin{center}
		\includegraphics[width=0.6\textwidth,clip,trim=10 30 50 35]{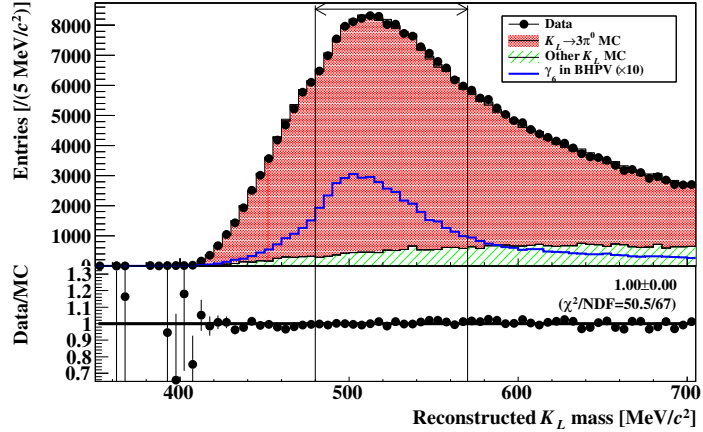}
		%trim = left bottom right top 
       \end{center}
	\caption{Invariant mass distribution of the $5 \gamma$ events (with the forward solution).
	The selection on $M_{K_{L}}^{\mathrm{backward}}$ is applied.
	In this plot, the $K_{L}\to 3 \pi^{0}$ MC (histogram in red) includes the subsequent $\pi^{0}\to e^{+}e^{-}\gamma$ decays.
        The blue histogram is for the events with the correct combination and $\gamma_{6}$ going into the BHPV detector
        in the simulation, and scaled by 10.
        The lines and arrow indicate the selection region.}
	\label{fig:5gamma-invariant mass}
\end{figure}

We required $480<M_{K_{L}}^{\mathrm{forward}}<570\,\mathrm{MeV}/c^{2}$,
because a simulation study showed that the forward solution was correct for most of the cases
in which the $\gamma_{6}$ hit the BHPV detector.
Events with $480<M_{K_{L}}^{\mathrm{backward}}<525\,\mathrm{MeV}/c^{2}$ were rejected
to reduce unnecessary backward solution events.
Comparison of the $M_{K_{L}}^{\mathrm{forward}}$ distribution between the data and MC is shown in Fig.~\ref{fig:5gamma-invariant mass}.
The MC result was normalized with the number of events after the cuts on the reconstructed $K_{L}$ mass.
The MC well reproduced the data,
though the distribution had no clear peak around the nominal $K_{L}$ mass due to events with incorrect photon combinations.
When we selected the events with proper photon combinations and $\gamma_{6}$ going into the BHPV in the simulation,
the distribution had a peak around the nominal $K_{L}$ mass
as the blue histogram in Fig.~\ref{fig:5gamma-invariant mass}.
The background contamination, 
mainly from the $K_{L} \to \pi^{+}\pi^{-}\pi^{0}$
and $K_{L} \to 3\pi^{0}$ decays with the subsequent Dalitz decay ($\pi^{0} \to e^{+}e^{-}\gamma$), 
was estimated to be 9.8\%. 

As a reference, events where all the six photons from the $K_{L} \to 3 \pi^{0}$ decay hit the calorimeter were reconstructed.
These events were called the ``6$\gamma$'' events, and the procedure was almost the same with that in 5$\gamma$ events
except the assumption that all 3$\pi^{0}$ came from the common decay vertex.
With the same normalization factor as in Fig.~\ref{fig:5gamma-invariant mass},
the number of the 6$\gamma$ events was found to be consistent between the data and MC
within the statistical uncertainty of 0.5\%.

%++++ figures 11 +++++++++++++++++++++++++++++++++++++++
%++++ figures 11 +++++++++++++++++++++++++++++++++++++++

%++++ figures 12 +++++++++++++++++++++++++++++++++++++++
\begin{figure}[h]
\begin{minipage}[t]{7.5cm}
	\begin{center}
		\includegraphics[width=\textwidth,clip,trim=5 25 45 35]{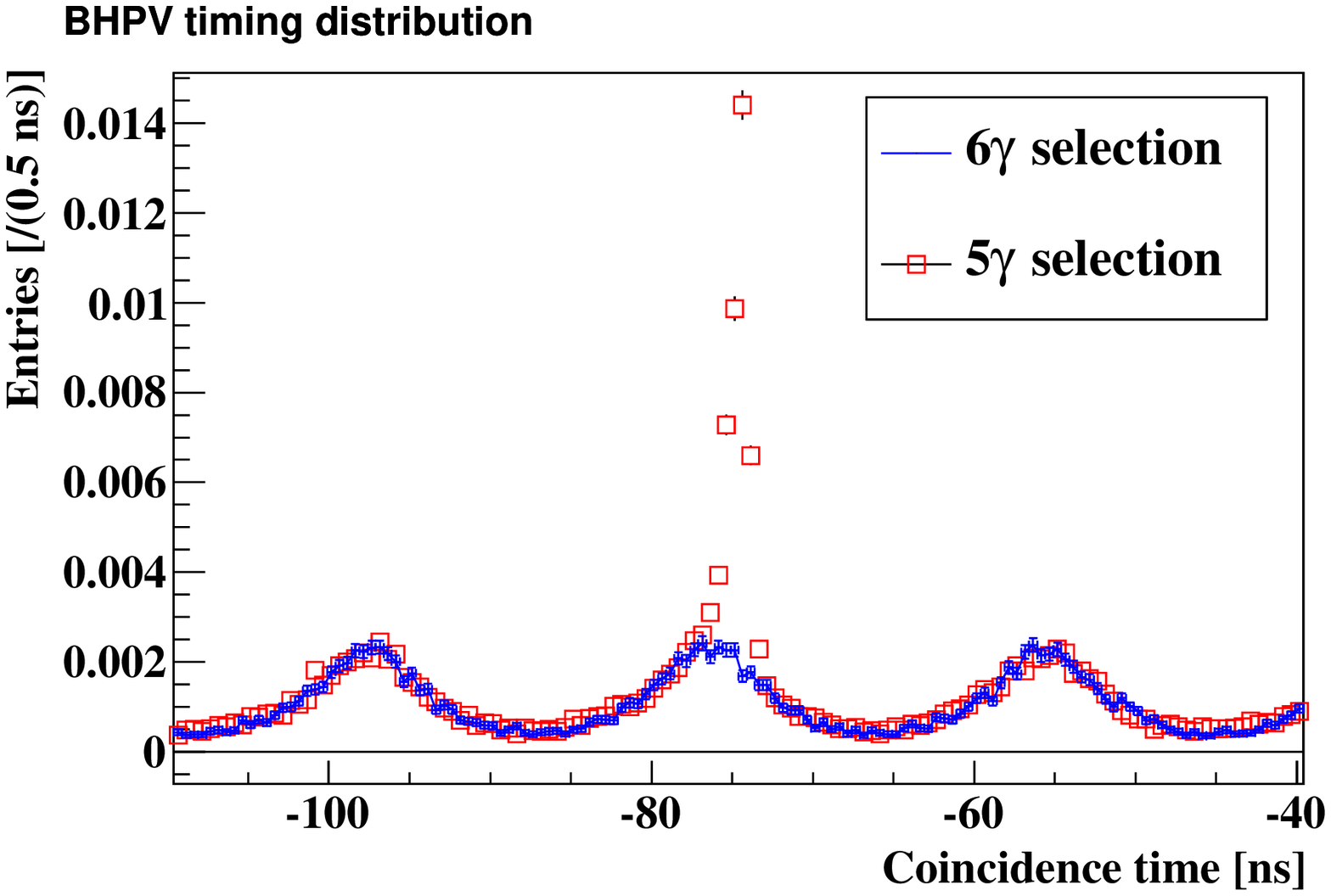}
        \end{center}
\end{minipage}
\begin{minipage}{0.5cm}$\:$ \end{minipage}
\begin{minipage}[t]{7.5cm}
	\begin{center}
		\includegraphics[width=\textwidth,clip,trim=5 5 45 35]{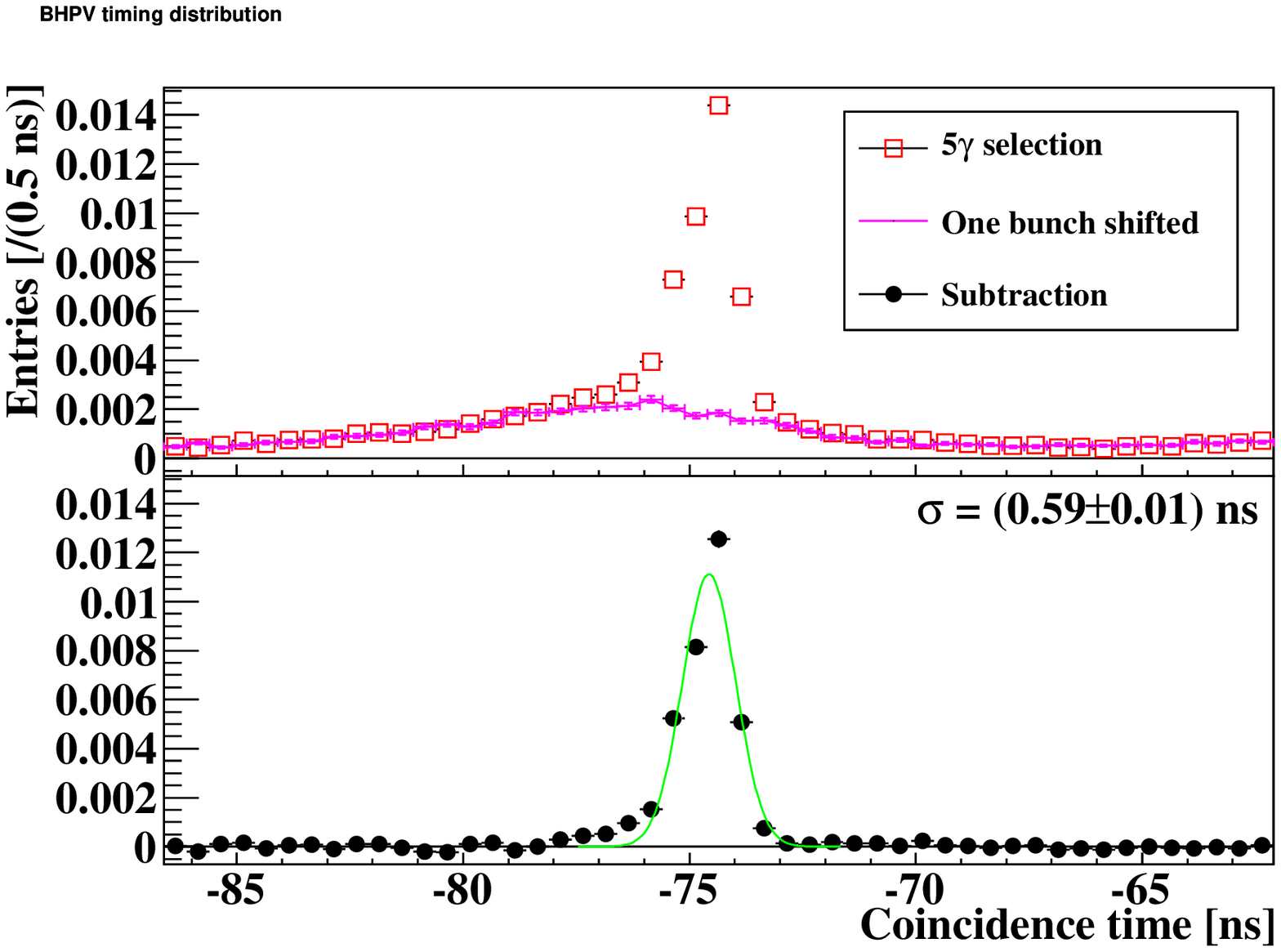}
        \end{center}
\end{minipage}
	\caption{(Left) Timing distribution of the BHPV hit events
	for the $5 \gamma$ (red) and $6 \gamma$ (blue) selections. 
        (Right top) Timing distribution of the $5 \gamma$ events between -86.6 and -62.6~ns (red)
        and the events in the previous bunch (between -107.8 and -83.7~ns) shifted by the cycle of 21.1~ns (purple).
        (Right bottom) Time distribution of the 5$\gamma$ events
        after subtracting the distribution of accidentals in the previous bunch.
	The line is a fit result by gaussian and the obtained $\sigma$ value is written in the right corner.
        }
        \label{fig:5gamma-timing distribution}
\end{figure}
%++++ figures 12 +++++++++++++++++++++++++++++++++++++++

%++++ figures 13 +++++++++++++++++++++++++++++++++++++++
\begin{figure}[h]
\begin{minipage}[t]{7.5cm}
	\begin{center}
		\includegraphics[width=\textwidth,clip,trim=5 0 45 35]{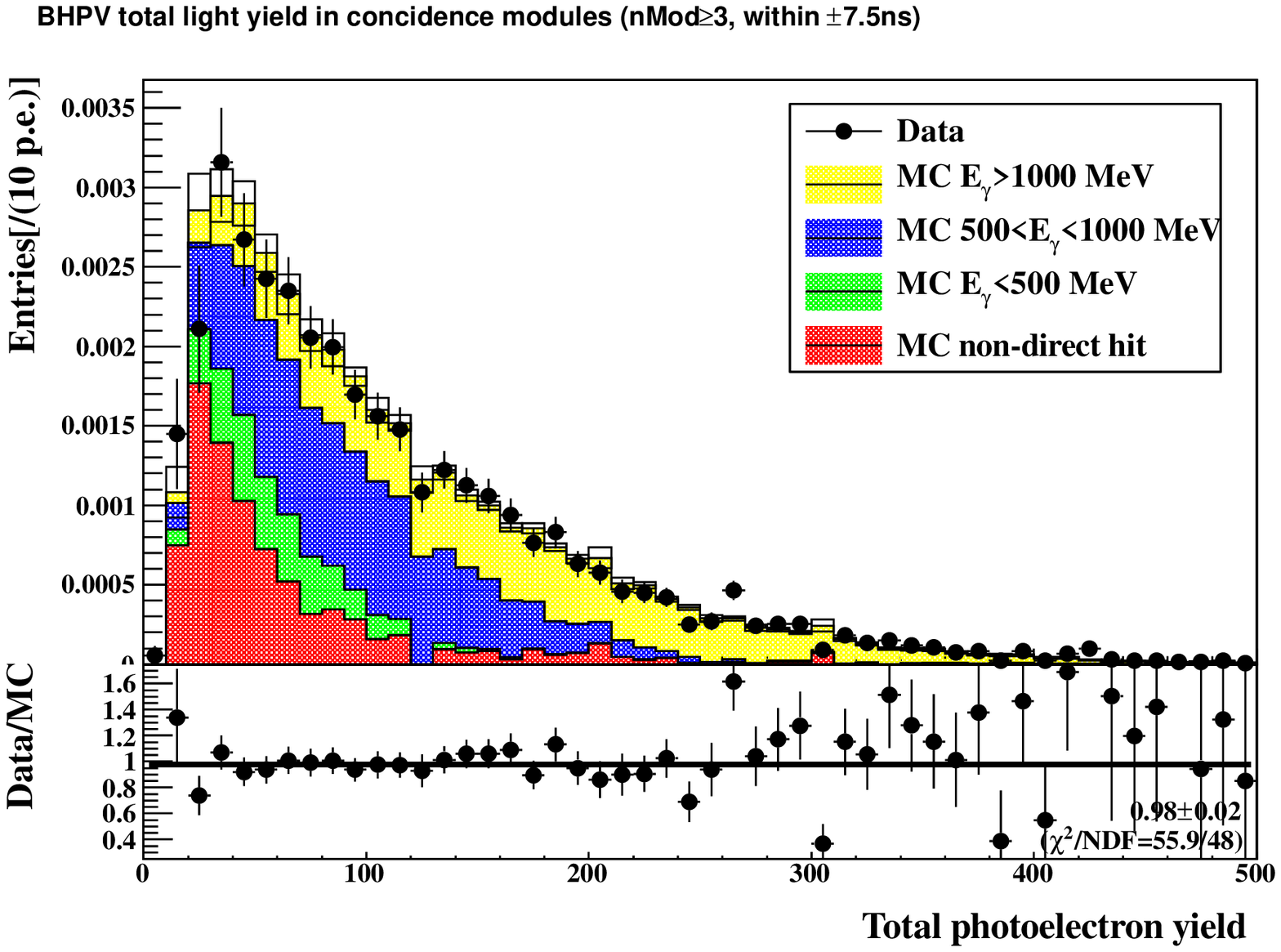}
         \end{center}
\end{minipage}
\begin{minipage}{0.5cm}$\:$ \end{minipage}
\begin{minipage}[t]{7.5cm}
	\begin{center}
		\includegraphics[width=\textwidth,clip,trim=30 35 40 35]{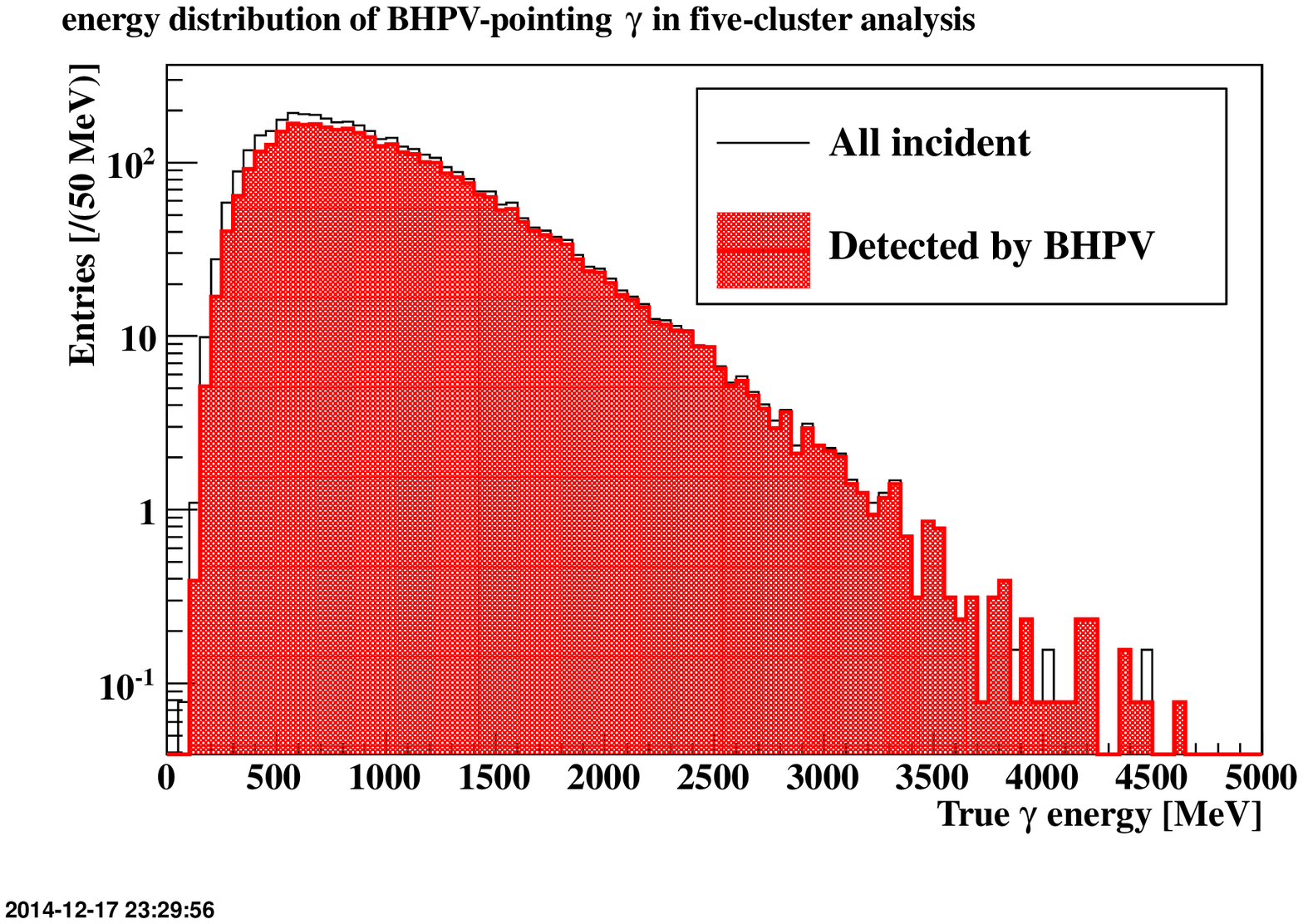}
		%trim = left bottom right top 
        \end{center}
\end{minipage}
	\caption{(Left top) Distribution of the total photoelectron yield.
        The solid points show the real data while the histograms show the MC simulation result.
        Energies of the missing $\gamma$ in the MC are classified by colors:
        0-500~MeV (green),  500-1000~MeV (blue), and~1000 MeV or more (yellow), and other events (red) 
        in which photon conversion points are outside of the BHPV detector (non-direct hits).
        The open box on each bin indicates the statistical error of the MC data.
        (Left bottom) Ratio of the real data to the MC data.	
        (Right) Distribution of the $\gamma_{6}$ energy in the MC simulation.}
        \label{fig:Gamma identification-photoelectron distribution}
\end{figure}
%++++ figures 13 +++++++++++++++++++++++++++++++++++++++

%++++ figures 14 +++++++++++++++++++++++++++++++++++++++
\begin{figure}[h]
	\begin{center}
		\includegraphics[width=0.6\textwidth,clip,trim=5 0 45 35]{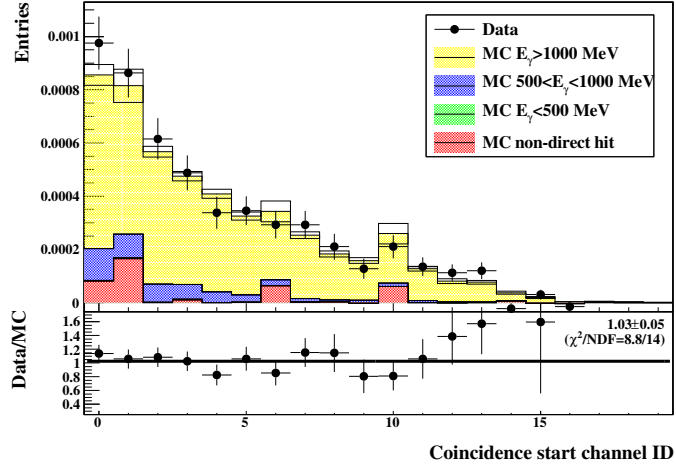}		
        \end{center}
	\caption{(Top) Distribution of the shower starting module in the events
	with the total photoelectron yield larger than 200.
        The solid points show the real data while the hatched histograms show the MC simulation result.
        See Fig.~\ref{fig:Gamma identification-photoelectron distribution} for the color codes.
        (Bottom) Ratio of the real data to the MC data.}
	\label{fig:Start-module-distribution}
\end{figure}
%++++ figures 14 +++++++++++++++++++++++++++++++++++++++

%------------------%
\subsection{BHPV photon response}
%------------------%
Now we examine the response of the BHPV detector using the $5\gamma$ and 6$\gamma$ events.
A photon hit in the detector was identified as three or more consecutive modules
with the outputs exceeding the 2.5~p.e. threshold in either or both of the left and right PMTs. 
Figure~\ref{fig:5gamma-timing distribution} shows the timing distribution of the photon hits in the BHPV detector
with respect to the timing determined by the CsI calorimeter\footnote{
The BHPV hit timing was defined as the hit times averaged over the modules
with hits after correcting time of propagation of the shower particles module by module.
The CsI hit timing was the weighted average of the photon hit times where the weight is given by the photon energy. 
Correction due to the time-of-flight between a decay vertex to hit positions was applied.}.
For the $6\gamma$ events, shown with the blue bars in Fig.~\ref{fig:5gamma-timing distribution} (left), 
there should be only accidental hits in the BHPV detector.
The periodic distribution seen in these hits reflects the beam bunch structure
in the slowly-extracted beam from the J-PARC MR.
On the other hand, for the $5\gamma$ events shown by the red open squares, 
a sharp peak is observed on top of the accidental hits.
Figure~\ref{fig:5gamma-timing distribution} (right bottom) shows the timing distribution of the $5\gamma$ events
after subtracting the distribution of the accidental hits in  Fig.~\ref{fig:5gamma-timing distribution} (right top).
The $\sigma$ value from the gaussian fit was found to be $0.59$~ns.
This is a clear evidence that the missing photons tagged by the $5\gamma$ events were successfully detected by the BHPV detector.

We compared the number of events within $\pm$7.5~ns of the peak in the timing distribution of the 5$\gamma$ events
between the data and MC.
Figure~\ref{fig:Gamma identification-photoelectron distribution} (left) shows the distribution 
of the total photoelectron yields observed by the hit modules in these events.
The total photoelectron yield was obtained by summing the outputs over the modules which recorded hits\footnote{
The output from a module was defined as the sum of the outputs exceeding 2.5~p.e. in the left and right PMTs.}
in three or more consecutive modules.
For the the data and MC, distributions of the accidental hits were subtracted.
The data and MC distributions agree.
As expected, the photoelectron yields increase with the energies of the $\gamma_{6}$ (the missing photon).
The energy distribution of the $\gamma_{6}$ going into the BHPV direction in the simulation is shown in Fig.~\ref{fig:Gamma identification-photoelectron distribution} (right).
We now focus on the events with the total photoelectron yield greater than 200.
The $N_{\mathrm{data}}^{>200\mathrm{p.e.}}$ and $N_{\mathrm{MC}}^{>200\mathrm{p.e.}}$ were defined
as the numbers of such events in the real data and the MC simulation, respectively.
In addition, we define $\eta=N_{\mathrm{data}}^{>200\mathrm{p.e.}}/N_{\mathrm{MC}}^{>200\mathrm{p.e.}}$.
Since the MC simulation shows that these events are mainly from the $\gamma_{6}$ with $>\! 1000$~MeV hitting the detector (90.3\%), 
$\eta$ is a good measure of the detector response to high energy photon.
If the detector works as expected in the MC simulation, $\eta$ gets close to 1.
Based on Fig.~\ref{fig:Gamma identification-photoelectron distribution},
$\eta=1.025\pm0.050\pm0.068$,
where the first and second errors represent the statistical and the systematic uncertainties, respectively.
A summary of the systematic errors is presented in Table~\ref{tab_SystError}.
The MC reproducibility of the BHPV was evaluated
by comparing the efficiency of each selection cut related to the BHPV between the data and MC.
For the other error sources, each condition was shifted within its uncertainty in the MC simulations
and changes of the event ratio with the $\gamma_{6}$ going to the BHPV were considered as the error.

\begin{table}[t]
	\caption{List of systematic uncertainties.}
	\label{tab_SystError}
	\begin{center}
		\begin{tabular}{lc}
			\hline\hline
			Error source & Relative error [\%] \\ \hline
			MC reproducibility on BHPV &  3.80 \\		%3.800
			$K_{L}$ momentum spectrum & $^{+1.70}_{-1.35}$ \\
			Beam position &  0.15 \\	%0.145
			Calorimeter energy resolution & 0.49 \\ 	%0.489
			Calorimeter position resolution & 4.98 \\	%4.975
			Detector alignment & 1.50 \\	\hline %1.503
			Total & $^{+6.68}_{-6.60}$ \\  \hline\hline
		\end{tabular}
	\end{center}
\end{table}

Finally, the detection efficiency for high energy photons was estimated from the $\eta$ value obtained above.
The efficiency was defined as
\begin{equation}
\epsilon_{\mathrm{data}}^{>1\mathrm{GeV}} = N_{\mathrm{data}}^{>1\mathrm{GeV}}
/ N_{\mathrm{true,incident}}^{>1\mathrm{GeV}},
\label{eq:EfficiencyDef}
\end{equation}
where $N_{\mathrm{data}}^{>1\mathrm{GeV}}$ and $N_{\mathrm{true,incident}}^{>1\mathrm{GeV}}$ were
the numbers of events with the $\gamma_{6}$ detected and incident in the BHPV detector, respectively,
when its energies was larger than 1~GeV.
We assumed
\begin{equation}
\frac{N_{\mathrm{data}}^{>200\mathrm{p.e.}}}{N_{\mathrm{data}}^{>\mathrm{1GeV}}}\simeq
\frac{N_{\mathrm{MC}}^{>200\mathrm{p.e.}}}{N_{\mathrm{MC}}^{>\mathrm{1GeV}}},
\end{equation}
and evaluated $N_{\mathrm{true,incident}}^{>1\mathrm{GeV}}$ with the simulation
($N_{\mathrm{MC,incident}}^{>1\mathrm{GeV}}$).
This assumption allows Eq.~(\ref{eq:EfficiencyDef}) to be deformed as follows:
\begin{eqnarray}
\epsilon_{\mathrm{data}}^{>1\mathrm{GeV}} & \simeq & ( N_{\mathrm{MC}}^{>1\mathrm{GeV}} / N_{\mathrm{MC}}^{>200\mathrm{p.e.}} \times N_{\mathrm{data}}^{>200\mathrm{p.e.}} )/ N_{\mathrm{MC,incident}}^{>1\mathrm{GeV}} \nonumber \\
	& = & \epsilon_{\mathrm{MC}}^{>1\mathrm{GeV}} \times \eta \label{eq:EffEq}
\end{eqnarray}
The efficiency in the MC simulation, written as $\epsilon_{\mathrm{MC}}^{>1\mathrm{GeV}}$,
was calculated to be $0.938\pm0.002(\mathrm{stat.})$.
Here, the inefficiency of $\sim$6\% mainly came from lack of the total radiation length
and would be reduced by adding modules to have enough thickness of the lead converter.
From Eq.~(\ref{eq:EffEq}), the efficiency for high energy photons was obtained as $\epsilon_{\mathrm{data}}^{\mathrm{>1GeV}} = 0.962\pm0.046(\mathrm{stat.})^{+0.064}_{-0.063}(\mathrm{syst.})$.
%%0.96172 = (0.937963+/-0.00169) * (1.02533+/-0.04951)
%%data yield = (511.774+/-22.4118)*1e-5*133262
%%MC yield = (499.13+/-10.154)*1e-5*132896

The $\eta$ value, which indicates the reproducibility of MC, is consistent with 1 within the error.
We concluded that high energy photons are successfully detected by this detector as expected
even when it was placed in the intense neutral beam.
The distribution of the shower-starting module between the data and MC in Fig.~\ref{fig:Start-module-distribution}
confirms further the detector performance.
The obtained efficiency, which is close to 1, indicates the excellent performance of the system as a photon veto detector.

%------------------------------%
\section{Summary}
%------------------------------%
In this paper, we have described a novel photon detector used in an intense neutral kaon beam line.
The aerogel Cherenkov radiation is adopted for the detection of electromagnetic showers,
and blindness to neutrons is expected.
According to MC simulations, which have been validated by tests experiments with positrons and protons,
efficiencies to photons with the energy larger than 1~GeV and to neutrons with the momentum of 2~GeV/$c$
are \textgreater 99.5\% and \textless 1\%, respectively.
The detector was partially installed in the first physics run of the KOTO experiment,
and performance to high energy photons was evaluated by tagging $K_{L}\to3\pi^{0}$ decay events.
It was confirmed the photon detection efficiency expected by the MC simulations was successfully achieved
within the 8.2\% uncertainty.
%We believe that the photon detector of this kind is an indispensable tool for $K_{L}$ rare decay experiments.

\section*{Acknowledgment}
We would like to express our gratitude to staffs of the KEK Proton Synchrotron and the Laboratory of Nuclear Science
for their cooperation in the test beam experiments.
We also thank the staff members of the J-PARC accelerator, Hadron Beam groups and the KEK Computing Research Center
for their supports in taking and analyzing the physics data.
Part of this work was supported by MEXT KAKENHI Grant Numbers 23224007, 18071006, 14046220 
and the Japan/US Cooperation Program.
Some of the authors were supported by Grant-in-Aid for JSPS Fellows.
%23224007*** : kiban S (since 2011~2016)
%18071007 : tokutei 2006-2011 New Developments of Flavor Physics
%18071006*** : tokutei 2006-2011 Search for flavor mixing and new physics with K_L→π^0νν^^- decay
%14046220*** : tokutei 2002-2005 (by Sugimoto, KEK) K+ and KL0 rare decay
%14046211 : tokutei 2002-2005 (by Yamanaka, Osaka) KLpi0nunu
%13047101 : tokutei 2001-2016 (by Kim, Tsukuba) Study of Mass Origin and Supersymmetry Physics
%11J00717 : my JSPS

\appendix

%------------------------- performance tests ---------------------------%
\section*{Appendix: Aerogel transmittance measurement}  
%------------------------- performance tests ---------------------------%
The transmittance of aerogel is known to be influenced mainly by two effects;
absorption and Rayleigh scattering.
According to \cite{Aschenauer}, it can be represented as
\addtocounter{equation}{5}
\begin{equation}
	T(\lambda)=\exp \left( -A^{\prime}\Delta x \right)\exp \left(- \frac{C \Delta x}{\lambda^4} \right),
        \label{eq:transmittance}
\end{equation}
where $\Delta x$ and $\lambda$ denote the thickness of aerogel and the wavelength of light, respectively.
$A^{\prime}$ and $C$ are constants.
The first (second) exponential represents the absorption (Rayleigh scattering) effect, 
and $A^{\prime}$ and $C$ characterize aerogel's transmittance property. 

These constants were measured using the setup shown in Fig.~\ref{fig:Aerogel transmission measurement setup}.
There were five LEDs with different colors, which were irradiated onto the aerogel sample under test 
through 2-mm-diameter holes.
The LEDs and aerogel sample were placed on movable tables controlled by a computer.
The transmittance was obtained by comparing light outputs from the main PMT behind the aerogel sample 
and those without the sample.
The stability of the LEDs was monitored by a separate PMT placed near the LEDs.
An example of the measurement results is shown in Fig.~\ref{fig:Aerogel transmission measurement result}
together with a fit result with Eq.~(\ref{eq:transmittance}).
In Table~\ref{table:transmittance}, we list the parameters obtained for the type-A and type-M aerogels averaged over many measurements and samples.
Combining with the information from the photoelectron calibration experiment,
these parameter measurements provide inputs to the simulation studies.
\addtocounter{figure}{14}
%++++ figures 15 +++++++++++++++++++++++++++++++++++++++
\begin{figure}[h]
\begin{minipage}{7.5cm}
	\begin{center}
		\includegraphics[width=\textwidth,bb=0 0 303 266]{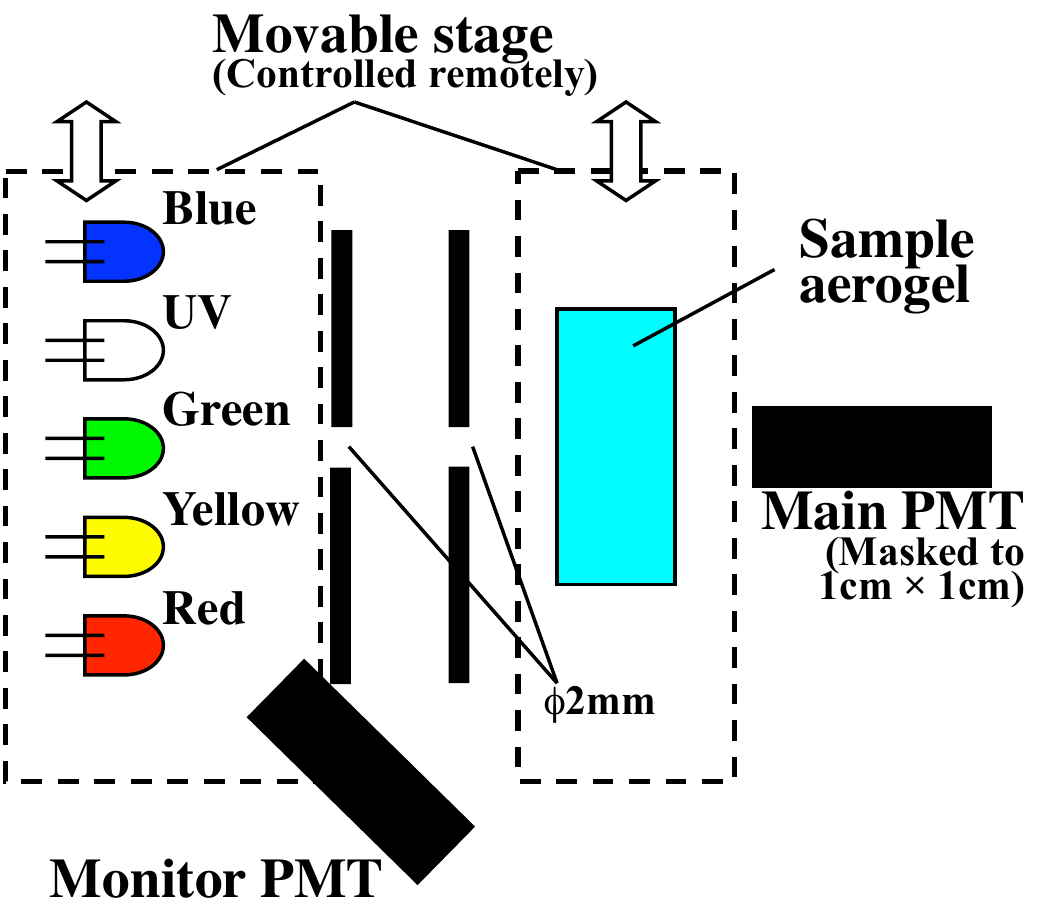}
        \end{center}
	\caption{Layout of the aerogel transmission measurement.
	The section inside the broken lines can be moved remotely.}
	\label{fig:Aerogel transmission measurement setup}
\end{minipage}
%++++ figures 15 +++++++++++++++++++++++++++++++++++++++
\begin{minipage}{0.5cm}$\:$ \end{minipage}
%++++ figures 16 +++++++++++++++++++++++++++++++++++++++
\begin{minipage}{7.5cm}
	\begin{center}
		\includegraphics[width=\textwidth,bb=0 0 400 339]{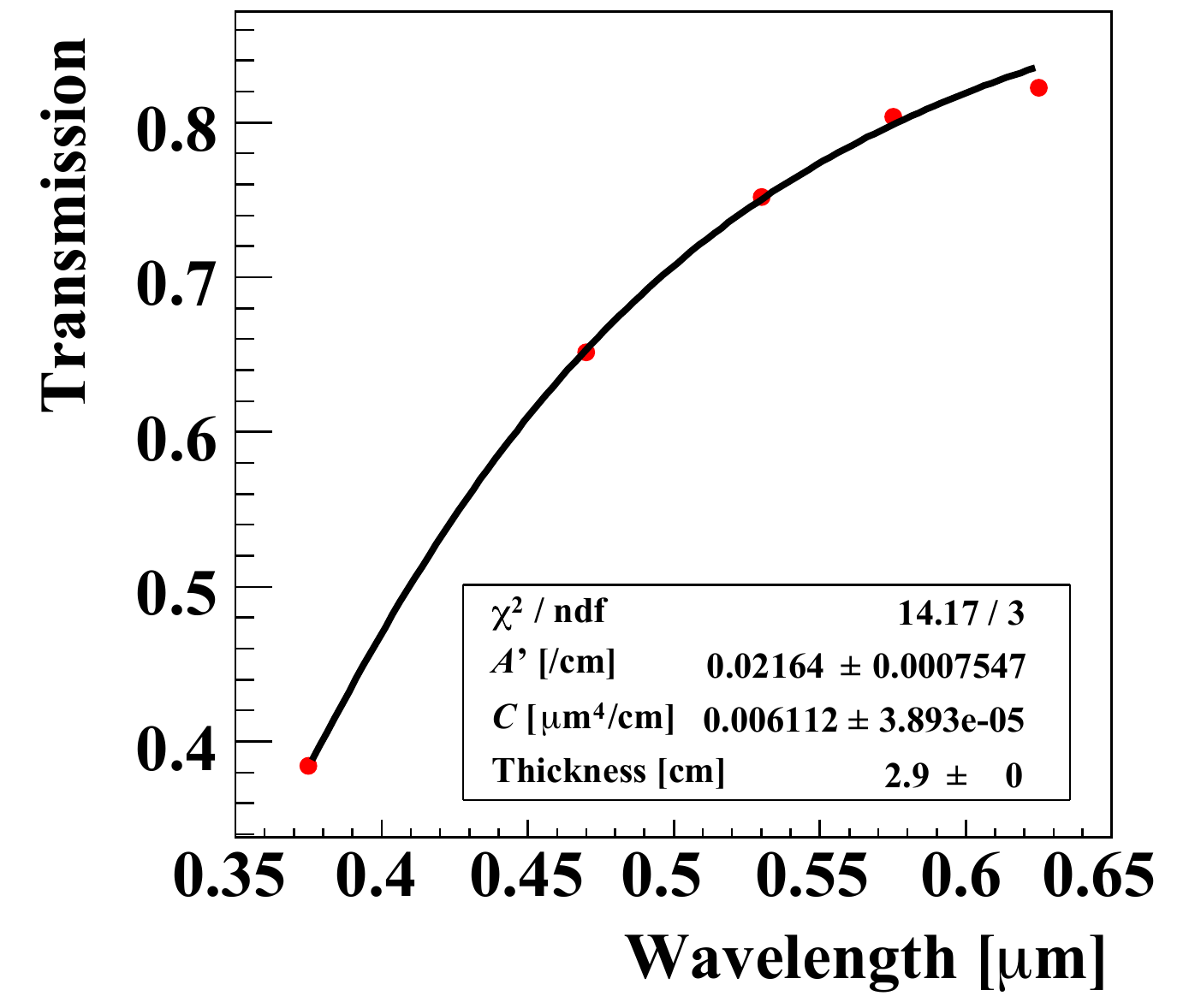}
        \end{center}
	\caption{Results of the aerogel transmission measurement.}
	\label{fig:Aerogel transmission measurement result}
\end{minipage}
\end{figure}
%++++ figures 16 +++++++++++++++++++++++++++++++++++++++

\addtocounter{table}{2}
%---------------------Table-------------------------%
\begin{table}[t]
\begin{center}
\caption{Summary of transmittance measurements. }
\begin{tabular}{ccc}
\hline\hline
Type    & $A$  &  $C$    \\ 
        &      &    [cm/$\mu$m$^4$]    \\ \hline
M & 0.96 & 0.0040   \\
A & 0.972 & 0.00692  \\ \hline\hline
\end{tabular}
\label{table:transmittance}
\end{center}
Note: The parameter $A$ is defined as $A=\exp (-A^{\prime}\Delta x)$, where $\Delta x$ is taken to be 1 cm.
\end{table}
%---------------------Table-------------------------%

\newpage

\bibliographystyle{ptephy}
%\bibliography{sample}
%
% once the .bbl file has been generated then place the text in your article.

\end{document}